\documentclass[twocolumn,npa,showpacs]{revtex4}
\usepackage{amsmath,bm}%
\usepackage{graphicx}%
\usepackage{ulem}

\begin{document}

\draft

\title{ $\Delta\Delta$ dibaryon structure in extended chiral
SU(3) quark model \footnote{The project is supported by Scientific
Research Foundation of Liaoning Education Department (No.
202122028)}}

\author{\footnotesize DAI Lian-Rong\footnote{Email
address:~dailr@lnnu.edu.cn\\
}}

\address{Department of Physics, Liaoning Normal University, 116029,
Dalian, P. R. China}

%\title{$\Delta\Delta$ dibaryon structure in extended chiral
%SU(3) quark model}
%\thanks{Project supported by the Scientific
%Research Foundation of Liaoning Education Department (No.
%202122028) }
%\author{L. R. Dai}\thanks{Email: dailr@lnnu.edu.cn,
%Tel:0411-84259629}
%\affiliation{Department of Physics, Liaoning
%Normal University, 116029, Dalian, P. R. China}

\begin{abstract} \textnormal{\small {The structure of
$\Delta\Delta$ dibaryon  is studied in the extended chiral SU(3)
quark model in which  vector meson exchanges are included. The
effect from the vector meson fields is very similar to that from
the one-gluon exchange (OGE) interaction. Both in the  chiral
SU(3) quark model and in the extended chiral SU(3) quark model,
the resultant mass of the $\Delta\Delta$ dibaryon  is lower than
the threshold of the $\Delta\Delta$ channel  but higher than that
of the $\Delta N\pi$ channel. }}
\end{abstract}


\maketitle

Recently the chiral $SU(3)$ quark model~[1] has been extended to
include vector chiral fields ~[2-4]. Using this extended chiral
$SU(3)$ quark model,the deuteron structure and $N-N$ scattering
process were studied. In this extended chiral $SU(3)$ quark model,
instead of the one gluon exchange (OGE) interaction, the vector
meson exchanges  play the dominate role in the short range part of
the quark-quark interactions. Since dibaryon systems have small
size, the short range behavior of the interaction must be
important for the dibaryon structures~[5,6]. In the work of
Glozman et al [7,8], the vector meson coupling was also included
to replace the OGE. It was pointed out that the spin-flavor
interaction is important in explaining the energy of the Roper
resonance, and a comparatively good fit to the baryon spectra was
obtained.

Many works show that the $\Delta \Delta$ dibaryon (deltaron;
$S=3,J^{\pi}=3^+$ and $T=0$, where $S$ is the spin, $J$ the total
angular momentum and $T$ the isospin) is an interesting candidate
of a nonstrange dibaryon. In 1987, Yazaki~[9] analyzed systems
with two nonstrange baryons in the framework of the cluster model
where the OGE interaction and the confining potential between two
quarks were considered. The result showed that among the $NN$,
$N\Delta$, and $\Delta\Delta$ systems, the $\Delta\Delta$ system
(deltaron with $S=3,T=0$) is the only system in which color
magnetic interaction (CMI) between two clusters is attractive.
Since the $\Delta$ is a resonance with a quite wide width and
easily decays into $N\pi$, the deltaron might have a large width
so that it cannot easily be detected in the experiment, even
though it is a bound state of $\Delta\Delta$, except that its mass
is below the threshold of the $NN\pi\pi$ channel. Wang et al.~[10]
studied the structure of the deltaron in terms of the quark
delocalization model. They found that the deltaron is a deeply
bound state with a binding energy of 320-390 MeV, namely, its
energy level is below the threshold of the $NN\pi\pi$ channel.
Yuan and Zhang et al.~[11] also studied the structure of the
deltaron in the chiral $SU(3)$ quark model, in which the OGE plays
the dominate role in the short range part of the quark-quark
interactions. They found that the resultant mass of the Deltaron
is lower than the threshold of the $\Delta\Delta$ channel but
higher than that of the $\Delta N\pi$ channel.

In this paper, we study the structure of deltaron in the extended
chiral $SU(3)$ quark model. The resonating group method (RGM) by
solving a coupled-channel equation is used, where the
$\Delta\Delta$ and $CC$ (hidden color) channels are all included.
In Sec. II, a brief introduction of the extended chiral $SU(3)$
quark model is outlined. The result is presented and discussed in
Sec. III. A conclusion is drawn in Sec. IV.

In the extended chiral $SU(3)$ quark model, besides the nonet
pseudo-scalar meson fields and the nonet scalar meson fields, the
coupling between vector meson fields and quarks is also
considered. With this generalization, the Hamiltonian of the
system can be written as
\begin{eqnarray}
& H & =\sum\limits_{i}T_i-T_{\rm G}+\sum\limits_{i<j}V_{ij},
\end{eqnarray}
and
\begin{eqnarray}
& V_{ij} & =V_{ij}^{\rm conf}+V_{ij}^{\rm OGE}+V_{ij}^{\rm ch},
\end{eqnarray}

\begin{eqnarray}
& V_{ij}^{\rm ch} & = \sum^{8}_{a=0} V^{{\rm s}_a} (\vec{r}_{ij})
+ \sum^{8}_{a=0}
   V^{{\rm ps}_a} (\vec{r}_{ij})+ \sum^{8}_{a=0}V^{{\rm v}_a} (\vec{r}_{ij})~
\end{eqnarray}

where $\sum\limits_{i}T_i-T_{\rm G}$ is the kinetic energy of the
system, and $V_{ij}$ includes all the interactions between two
quarks. $V_{ij}^{\rm conf}$ is the confinement potential taken as
quadratic form, $V_{ij}^{\rm OGE}$ is the OGE interaction, and
$V_{ij}^{\rm ch}$ represents the interactions from the chiral
field coupling, which, in the extended chiral $SU(3)$ quark model,
includes the scalar meson exchange $V_{ij}^{\rm s}$, the
pseudo-scalar meson exchange $V_{ij}^{\rm ps}$ and the vector
meson exchange $V_{ij}^{\rm v}$ potentials. In Eq.(2), the OGE is
taken in the usual form ~[2],while the confinement potential is
chosen in the quadratic form

\begin{eqnarray}
V_{ij}^{\rm
conf}=-\lambda_{i}^{c}\cdot\lambda_{j}^{c}~a_{ij}^{c0}-\lambda_{i}^{c}\cdot\lambda_{j}^{c}a_{ij}^{c}r_{ij}^2~.
\end{eqnarray}

The quark-chiral field (scalar,pseudo-scalar and vector nonet
mesons) induced interactions are eqs. (5-7)

\begin{eqnarray}
V^{{\rm s}_a} (\vec{r}_{ij}) & = & -C(g_{\rm ch}, m_{s_a},
\Lambda_{c}) X_{1}(m_{s_a}, \Lambda_{c}, r_{ij})
\lambda_{a}(i)\lambda_a (j)\nonumber\\
 &&{} + V^{\vec{\ell} \cdot \vec{s}}_{s_a} (\vec{r}_{ij})\,\, ,
\end{eqnarray}

\begin{eqnarray}
V^{{\rm ps}_a}(\vec{r}_{ij}) & = & C(g_{\rm ch}, m_{{\rm ps}_a},
\Lambda_{c}) \frac{m_{{\rm ps}_a}^{2}}{12m_{qi}m_{qj}}\{
X_{2}(m_{{\rm ps}_a}, \Lambda_{c},
r_{ij})\nonumber\\
 &&{}
 (\vec{\sigma}_{i} \cdot \vec{\sigma}_{j})
  + \left ( H(m_{{\rm ps}_a} r_{ij}) - (\frac{\Lambda_{c}}{m_{{\rm ps}_a}} )^{3}
H(\Lambda_{c} r_{ij} ) \right )\nonumber\\
 &&{}
\hat{S}_{ij}\} \lambda_{a}(i) \lambda_{a}(j)\,\, ,
\end{eqnarray}

\begin{eqnarray}
 V^{{\rm v}_a} (\vec{r}_{ij}) & = & C(g_{\rm chv}, m_{v_a}, \Lambda_c)
X_{1}(m_{v_a}, \Lambda_c, r_{ij})\lambda_{a}(i)\lambda_a(j) \nonumber\\
  & + & C(g_{\rm chv}, m_{v_a}, \Lambda_c)
\frac{m_{v_a}^{2}}{6m_{qi}m_{qj}}(1+\frac{f_{\rm chv}}{g_{\rm
chv}} \frac{m_{qi}+m_{qj}}{M_p}\nonumber\\
 &&{}
+\frac{f_{\rm chv}^2}{g_{\rm chv}^2}\frac{m_{qi} m_{qj}}{M_p^2})
 \{  X_{2}(m_{v_a}, \Lambda_c,r_{ij})
(\vec{\sigma}_{i} \cdot \vec{\sigma}_{j})  \nonumber\\
  & - &\frac{1}{2} \left ( H(m_{v_a} r_{ij}) - (\frac{\Lambda_c}{m_{v_a}}
)^{3} H(\Lambda_c r_{ij} ) \right )\hat{S}_{ij} \}\nonumber\\
 &&{}
\lambda_{a}(i)\lambda_{a}(j) +  V^{\vec{\ell} \cdot \vec{s}}_{{\rm
v}_a} (\vec{r}_{ij}),
\end{eqnarray}
and $M_p$ is a mass scale, taken as proton mass. The detailed
formula expressions can be found in Ref. [1,2]. The coupling
constant of the OGE and the strength of confinement potential are
determined by the stability condition of N and mass difference
between $\Delta$ and $N$. The parameters $g_{\rm ch}$ is the
coupling constant for the scalar and pseudo-scalar chiral field
couplings, which can be determined from experimental value. For
vector meson nonet field coupling, the vector coupling and tensor
coupling constants $g_{\rm chv}$ and $f_{\rm chv}$  are taken to
be the same values as those used in the study of deuteron and $NN$
phase shift [2]. The meson masses ($m_{ps},~m_{s}$, and $m_{v}$)
are taken to be the experimental values.  Only the mass of
$\sigma$ meson ($m_{\sigma}$) is treated as an adjustable
parameter.  The cut-off mass $\Lambda$ is taken to be $1100$MeV
for all mesons. All parameters used here are shown in Table 1,
which are determined in the $NN$ scattering calculation by fitting
the binding energy of deuteron.
\begin{table}
\caption{  Model parameters and the the corresponding binding
energies $B_{\rm deu}$ of deuteron.}
\begin{small}
\begin {center}
\begin{tabular}{|c|c|c|c|}
\hline
%($~$)
                       & Chiral $SU(3)$&
\multicolumn{2}{|c|}{Extended  chiral $SU(3)$ }   \\
                       & quark model&
\multicolumn{2}{|c|}{quark model}   \\
\hline
                       &          & ~~~~~~ set I ~~~~~&  set
                        II
              \\ \hline
$b_u (fm)$             & 0.5      & ~~~~~~ 0.45  ~~~~   & 0.45      \\
$g_{NN\pi}$            & 13.67    & 13.67    & 13.67     \\
$g_{ch}$               & 2.621    & 2.621    & 2.621     \\
$g_{chv}$              & 0        & 2.351    & 1.972     \\
$f_{chv}/g_{chv}$      & 0        & 0        & 2/3       \\
$m_{\sigma}(MeV)$      & 595      & 535      & 547       \\
$g_u$                  & 0.886    & 0.293    & 0.399     \\
$\alpha_s (g_u^2)$     & 0.785    & 0.086    & 0.159     \\
$a_{uu}(MeV/fm^2)$     & 48.1     & 48.0     & 42.9      \\
\hline
                       &          &          &           \\
$B_{deu}(MeV)$         & 2.13     & 2.19     & 2.14      \\
\hline
\end{tabular}
\end{center}
%\end{flushleft}
\end{small}
\end{table}

\begin{table*}
\caption{Binding energy $B$ and rms $\overline{R}$ of the deltaron
$B=-(E_{\textrm{deltaron}}-2M_{\Delta})$,
$\overline{R}=\sqrt{\langle\,r^2 \rangle}$.}
\begin{small}
\begin {center}
\begin{tabular}{lllcccc}
\hline\hline &&&$\Delta\Delta(L=0)$&$\Delta\Delta\left
(\begin{array}{c}L=0\\+2\end{array} \right)$
&$\begin{array}{c}\Delta\Delta\\ CC\end{array}(L=0)$
&$\begin{array}{c}\Delta\Delta\\ CC\end{array}
\left(\begin{array}{c}L=0\\+2\end{array} \right)$\\
\hline
Chiral SU(3) quark model &&$B$(MeV)&23.0&29.1&40.6&48.3\\
&&$\overline{R}$(fm)&0.98&0.96&0.88&0.88\\
Extended chiral SU(3) quark model&~ set I~ &$B$(MeV)&56.1&62.4&78.2&84.0\\
&&$\overline{R}$(fm)&0.80&0.80&0.76&0.76\\
&~ set II~ &$B$(MeV)&41.7&48.0&64.7&70.4\\
&&$\overline{R}$(fm)&0.84&0.84&0.78&0.78\\
\hline\hline
\end{tabular}
\end{center}
%\end{flushleft}
\end{small}
\end{table*}

 We choose the two-cluster configuration as the dibaryon's model
space~ [11-13].  The $CC$ channel has the form
\begin{eqnarray}
|CC\rangle=-\frac{1}{2}|\Delta\Delta\rangle +\frac{\sqrt{5}}{2}
A_{STC}|\Delta\Delta\rangle
\end{eqnarray} where $A_{STC}$ stands for the
antisymmetrizer in the spin-isospin-color space.

 In our present calculation, the mixture of the $L=0$ and $L=2$ states which shows
the effects of the tensor forces in OGE and chiral field induced
potentials are also considered, namely, the
two-channel-four-state, $\Delta\Delta(L=0)$, $\Delta\Delta(L=2)$,
$CC(L=0)$, and $CC(L=2)$, calculation is performed~[11-14].

In the coupled-channel bound-state calculation, one must carefully
eliminate forbidden states, which may spoil the numerical
calculation. In the deltaron case, there exists a state, which has
the zero eigenvalue of the normalization operator $\langle N
\rangle=0$ due to the Pauli blocking effect. It reads
\begin{eqnarray}
|\Psi\rangle_{\textrm{forbidden}}=|\Delta\Delta\rangle-
\frac{1}{2}|CC\rangle\,\, .
\end{eqnarray}
Performing an off-shell transformation, this nonphysical degree of
freedom can be eliminated and the reliable result can be achieved.

By using the model parameters shown in Table I, the $NN$
scattering phase shifts and the  binding energy of deuteron can be
well reproduced~[2]. Here we use the same sets of parameters to
study the structure of deltaron. The calculated binding energies
of deltaron and the corresponding root-mean-square radius (rms) in
the extended chiral $SU(3)$ quark model are listed in Table II. In
order to compare the result with other model calculations, the
results of the  chiral $SU(3)$ quark model [1] are also given in
Table II. The results of the extended chiral $SU(3)$ quark model
for two different cases are shown. one is no tensor coupling of
the vector mesons with $f_{\rm chv}/g_{\rm chv}=0$ (set I),
another involves tensor coupling of the vector mesons with $f_{\rm
chv}/g_{\rm chv}=2/3$ (set II). The calculation is carried out in
four different combinations: $\Delta\Delta~(L=0)$,
$\Delta\Delta~(L=0$ and $2)$, $\Delta\Delta+CC~(L=0)$, and
$\Delta\Delta+CC$ ($L=0$ and $2$).

It is shown that the binding energy of the deltaron is indeed lower
than the threshold of the $\Delta\Delta$ channel, which is always
several tens MeV in all cases. Since the deltaron mass is still
higher than the mass of $N\Delta\pi$, the deltaron seems not to be a
narrow width dibaryon.

Our calculation shows that the channel coupling effect is much
larger than the $L$ state mixing effect, which are all caused by
the tensor interaction. The largest binding energy of deltaron
appears in the extended chiral SU(3) model in the case of set I.
It means that the vector chiral fields offer substantial
attractions across two $\Delta$ clusters, so that the deltaron
becomes more bound. The tensor coupling from vector fields in the
case of set II in the extended chiral SU(3) model will reduce
$\sim 14$ Mev in the deltaron binding energy as compared to that
in set I case.

It is shown from Table 1 that the coupling constant of the OGE is
greatly reduced when the vector meson field coupling is
considered. The coupling constant $\alpha_s$  of the OGE between
$u(d)$ quarks is determined by fitting the mass difference between
$\Delta$ and $N$. For both parameters shown in the set I and the
set II cases, $\alpha_s  < 0.2$ which is much smaller than the
value $(0.78)$ used in the  chiral $SU(3)$ quark model. The
results manifest that the OGE interaction is quite weak in the
extended chiral $SU(3)$ quark model. Instead of the OGE, the
vector meson exchanges play dominate role in the short range part
of the interaction between two quarks, so that the mechanism of
the quark-quark short range interaction of the two models is
totally different. The quark-quark short range interaction is from
the OGE in the  chiral $SU(3)$ quark model, while it is mainly
from vector meson exchanges in the extended chiral $SU(3)$ quark
model. Furthermore, The binding energy of deltaron in the extended
chiral $SU(3)$ quark model is quite similar as that of the
 chiral $SU(3)$ quark model. The results tell us that no
matter whether the OGE or the vector meson exchange controls the
quark-quark short range interaction, the main properties of
deltaron keep unaffected. In addition, it is also shown from the
calculation that the tensor coupling of the vector chiral field
reduces the binding energies of deltaron.

In summary, the vector meson exchange effect on deltaron in the
extended chiral SU(3) quark model is studied. The model space is
enlarged to include the CC channel. The contributions from various
chiral fields are included. The results show that the binding
energy of the deltaron is still ranged around several tens MeV
when the vector meson exchanges dominate the short range part of
the quark-quark interaction. Our results show that the mass of the
deltaron is always smaller than that of $\Delta\Delta$, but larger
than that of $\Delta N\pi$, which are quite similar to those of
the  chiral $SU(3)$ quark model.

\vspace{1.0cm} \noindent {\bf Acknowledgement}

 The author thanks
Prof. Zhang Zong-ye and Prof. Yu You-wen for their helpful
discussions.

\vspace{0.5cm} {\bf\noindent {\large References}}
\begin{small} \vskip .2cm
\noindent [1]
  Zhang Z Y, Yu Y W, Shen P N, Dai L R, Faessler A and
  Straub U 1997
 {\it Nucl. Phys.}\ A {\bf 625} 59\\
\noindent [2] Dai L R, Zhang Z Y, Yu Y W and Wang P 2003
  {\it Nucl.\ Phys.}  A {\bf 727} 321\\
\noindent [3] Zhang Z Y, Yu Y W, Wang P and Dai L R 2003
  {\it Commun.\ Theor.\ Phys.}  {\bf 39} 569\\
\noindent [4] Zhang Z Y, Yu Y W  and  Dai L R 2003
  {\it Commun.\ Theor.\ Phys.} {\bf 40} 332\\
\noindent [5] Jaffe R L 1977 {\it  Phys.\ Rev.\ Lett.} {\bf 38}
195\\
\noindent [6]
  Takeuchi S and Oka M 1991
 {\it Phys.\ Rev.\ Lett.} {\bf 66} 1271\\
\noindent [7] Glozman L Y and  Riska D O 1996 {\it Phys. Reports }
{\bf 268} 263 \\
\noindent [8] Glozman L Y  2000 {\it Nucl. Phys.} A {\bf 663} 103c
\\
\noindent [9]
  Yazaki K 1987 {\it Prog.\ Theor.\ Phys.\ Suppl.} {\bf 91} 146\\
\noindent [10]
  Wang F, Wu G H, Teng L J and Goldman T 1992
   {\it Phys.\ Rev.\ Lett.} \  {\bf 69} 2901; 1995 {\it  Phys.\ Rev.\ C }{\bf 51}
   3411; Pang H R, Ping J L, Chen L Z and Wang F  2004 {\it Chin. Phys. Lett.} {\bf
   21}
   1455; Lu X F, Ping J L and Wang F 2003 {\it  Chin. Phys. Lett.} {\bf
   20} 42\\
\noindent [11]
 Yuan X Q, Zhang Z Y, Yu  Y W  and Shen P N 1999
 {\it Phys.\ Rev.}\ C {\bf 60} 045203\\
\noindent [12]
 Yu Y W, Zhang Z Y and Yuan X Q 1999
 {\it Commun.\ Theor.\ Phys.}\  {\bf 31} 1\\
\noindent [13]
  Harvey  M 1981  {\it Nucl.\ Phys.}\ A {\bf 352} 326\\
\noindent [14]
  Zhang Z Y, Yu Y W, Shen P N, Shen X Y and Dong Y B
  1993 {\it Nucl.\ Phys.} A {\bf 561} 595\\
\end{small}
\end{document}